\documentclass[12pt]{article}
\usepackage[dvips]{graphicx}
\usepackage{latexsym}
\parskip=\medskipamount

\newcommand {\cD}{{\cal D}}

\newcommand {\cI}{{\cal I}}

\newcommand {\cN}{{\cal N}}

\newcommand {\cP}{{\cal P}}

\newcommand {\cV}{{\cal V}}

\def\a{\alpha}
\def \bi{\bibitem}

\def\b{\beta}

\def\d{\delta}

\def\g{\gamma}
\def\G{\Gamma}

\def\k{\kappa}
\def\l{\lambda}
\def\m{\mu}

\def\o{\omega}
\def\p{\pi}
\def\q{\theta}
\def\r{\rho}
\def\s{\sigma}
\def\t{\tau}

\def\z{\zeta}
\def\D{\Delta}
\def\F{\Phi}
\def\J{\Psi}
\def\L{\Lambda}
\def\O{\Omega}

\def\S{\Sigma}
\def\U{\Upsilon}

\newcommand{\ad}{{\dot{\alpha}}}                           
\newcommand{\bd}{{\dot{\beta}}}                            
\newcommand{\ve}{\varepsilon}                            

\newcommand{\pa}{\partial}                           
\newcommand{\hf}{\frac12}

\newcommand{\vf}{\varphi}
\newcommand{\sect}[1]{\setcounter{equation}{0}\section{#1}}

\newcommand{\be}{\begin{equation}}
\newcommand{\ee}{\end{equation}}
\newcommand{\bea}{\begin{eqnarray}}
\newcommand{\eea}{\end{eqnarray}}
\newcommand{\non}{\nonumber}
\begin{document}

\begin{titlepage}
\thispagestyle{empty}

\begin{flushright}
hep-th/0302205 \\
February, 2003 \\
Revised version: March, 2003
\end{flushright}
\vspace{5mm}

\begin{center}
{\Large \bf
On the Background Field Method Beyond One Loop: \\
A manifestly covariant derivative expansion \\
in super Yang-Mills theories }
\end{center}

\begin{center}
{\large S. M. Kuzenko and I. N. McArthur}\\
\vspace{2mm}

\footnotesize{
{\it
School of Physics, The University of Western Australia\\
Crawley, W.A. 6009, Australia}
} \\
{\tt  kuzenko@cyllene.uwa.edu.au},~
{\tt mcarthur@physics.uwa.edu.au}
\end{center}
\vspace{5mm}

\begin{abstract}
\baselineskip=14pt
There are currently many string inspired conjectures 
about the structure of the low-energy effective action for 
super Yang-Mills theories which require explicit multi-loop 
calculations. In this paper, we develop 
a manifestly covariant derivative expansion of superspace
heat kernels and present a scheme to evaluate multi-loop
contributions to the effective action in the framework 
of the background field method. The crucial ingredient 
of the construction is a detailed analysis of the properties
of the parallel displacement propagators associated
with Yang-Mills supermultiples in $\cN$-extended
superspace. 
\end{abstract}

\vfill
\end{titlepage}

\newpage
\setcounter{page}{1}

\renewcommand{\thefootnote}{\arabic{footnote}}
\setcounter{footnote}{0}
\sect{Introduction and outlook}

The background field method 
\cite{DeWitt67,Hon,tH1,Kallosh,AFS,tH2,DeWitt3,Ab,Boul,Hart}
is a powerful tool for the study of
quantum dynamics in gauge theories, including gravity.
Its main merit is a manifestly gauge invariant definition 
of the quantum effective action. 
Renormalization of the Yang-Mills theories 
in this approach is not much harder  
to establish \cite{Kallosh,KZ} than in the conventional one, 
and the $S$-matrix is correctly reproduced \cite{AGS,BC,FPQ}. 
The background field method is universal enough
to admit natural superspace generalizations; 
such extensions have been constructed for $\cN=1$
super Yang-Mills theories and supergravity  
\cite{GRS,GS,GGRS,GZ1,GZ2}
and for $\cN=2$ super Yang-Mills theories \cite{N=2BFM}
(see also \cite{HST} for earlier attempts to develop
a background field formulation for some $\cN=2$ SYM models,
including the $\cN=4$ SYM theory).

The price to pay for the manifest gauge invariance of 
the effective action is that one has to deal with propagators 
in arbitrary background fields, and these are impossible to 
compute exactly. At the one-loop level, this does not lead to  
any serious problems, since the famous Schwinger-DeWitt technique
\cite{Schw1,DeWitt1} and its numerous generalizations
\cite{BV,NSVZ,Nep,Avr1,Gus,MG,PB} allow one to compute 
the effective action to any given order in the derivative expansion.
Beyond the one-loop approximation, however, the full power of the 
background field method has never been exploited. 
A necessary precursor to such multi-loop calculations should be 
a manifestly gauge covariant derivative expansion 
of the Feynman propagators in the presence of arbitrary background
fields. This has essentially existed for non-supersymmetric theories 
since the mid 1980's \cite{BV,Avr1} (although, in our opinion,  
its significance has not yet been fully appreciated), 
while such an approach has never been elaborated for 
supersymmetric gauge theories in superspace.
In the non-supersymmetric case, 
a manifestly covariant calculus has only been worked out in detail 
for computing the UV divergences of the effective action \cite{JO}.
Otherwise (and often for counterterm calculations only), 
it has been usual to  resort to one of the following 
options\footnote{There also exists the so-called string-inspired 
approach to perturbative QFT, see \cite{Schubert} for a review and 
references.}
(see \cite{BV2,BvdV} for a more detailed discussion):
(i) a combination of conventional perturbation theory with 
the background field formalism, with manifest background gauge
invariance sacrificed at intermediate stages
\cite{Ab,CM,IO,GRS,GGRS,GZ2,GorS,PGJ};
(ii) the use of normal coordinates in curved space \cite{BP,BO}
or the Schwinger-Fock gauge for Yang-Mills fields
(see \cite{NSVZ} and reference therein), usually in conjunction 
with momentum space methods. 

One can hardly  overestimate the importance of developing
a manifestly gauge covariant derivative expansion of the {\it finite}
part of the (low energy)
effective action beyond the one-loop approximation\footnote{In this
paper we are interested in the local (or low energy) expansion of 
the effective action. Concerning the nonlocal part of the effective
action, the interested reader is referred to \cite{BM}
for a recent comprehensive discussion and references.}.
A great many research groups have contributed to solving 
the problem in the non-supersymmetric case, 
and it is simply impossible to give all references here. 
The present paper is aimed at completing this program
for the case of supersymmetric  Yang-Mills theories.
Crucial for our considerations were 
the ideas and techniques developed by
DeWitt \cite{DeWitt1} (in particular the concept of 
parallel displacement propagator), 
Barvinsky and Vilkovisky \cite{BV} (including 
the covariant Taylor expansion), as well as
Avramidi's extension  \cite{Avr1} 
of the covariant calculus introduced in \cite{BV}. 
Of course, in the beginning was Schwinger \cite{Schw1}.

This paper is organized as follows. In sect. 2 we develop
a manifestly covariant derivative expansion of the 
heat kernels in ordinary Yang-Mills theories, extending and 
generalizing previous results \cite{DeWitt1,BV,Avr1}.
In sect. 3 we outline a manifestly covariant scheme
for evaluating multi-loop diagrams in Yang-Mills theories
in the framework of the background field method. 
Sect. 4 is devoted to a detailed analysis of the properties
of the parallel displacement propagators associated 
with Yang-Mills supermultiplets in $\cN$-extended flat superspace. 
A superspace covariant Taylor expansion is also derived. 
The Fock-Schwinger gauge in $\cN$-extended superspace is 
described in detail for completeness\footnote{Unlike numerous 
previous quantum calculations with background fields, 
at {\it no} stage in our scheme 
is there need to use the Fock-Schwinger gauge.}.  
In sect. 5 we provide a manifestly covariant derivative expansion 
of the superfield heat kernels in $\cN=1$ super Yang-Mills theories. 
The exact superpropagator in a covariantly constant background 
is computed in sect. 6. Finally, the appendix contains 
the proof of a technical lemma.  

To keep  the background covariance of the effective action manifest, 
Barvinsky and Vilkovisky \cite{BV}, and more recently 
B\"ornsen and van de Ven \cite{BvdV}, eliminated the momentum 
representation from their consideration. Our approach differs 
conceptually from \cite{BV,BvdV} in that we {\it do} preserve 
the momentum representation in a manifestly covariant way.

The approach developed in this paper opens the way 
to computing low-energy effective actions in super Yang-Mills
theories beyond one loop. A supergravity extension is 
quite feasible, and old results on the proper-time technique
in curved $\cN=1$ superspace \cite{McA,BK2,BK} 
should be relevant. 
We believe that the results will also be helpful, for instance, 
for a better understanding  of (i) numerous non-renormalization
theorems which are predicted by the AdS/CFT conjecture
and relate to the explicit structure of the
low energy effective action in $\cN=4$ super Yang-Mills
theory (see \cite{CT,BKT} for more details
and references); (ii) quantum deformation of 
the superconformal symmetry in $\cN=2,4$ SCFTs \cite{KMT}
at higher loops.

\sect{Non-supersymmetric case}

We consider a Green's function,
$G^i{}_{i'} (x,x') = {\rm i}\, 
\langle \vf^i(x) \, {\bar \vf}_{i'} (x') \rangle $, 
associated with a quantum field $\vf$, which  transforms in some 
representation of the gauge group, and its conjugate $\bar \vf$. 
The Green's function satisfies the equation   
\be
\D_x \, G (x, x') = - \d^d (x-x') \, {\bf1} ~, 
\qquad 
\D = \nabla^m \nabla_m + \cP ~, \qquad
{\bf 1} = (\d^i{}_{i'})~,
\ee
with $\nabla_m = \pa_m +{\rm i} \,A_m $ the gauge covariant 
derivatives, $[ \nabla_m, \nabla_n ] = {\rm i}\, F_{mn}$, 
and $\cP(x) $ a local matrix function of the background field
containing a mass term $(-m^2 ){\bf 1}$. 
With respect to the gauge group, $\nabla_m$ and $\cP$ 
transform as follows
\be
\nabla_m ~\to~ {\rm e}^{{\rm i} \l(x)} \, \nabla_m\,
{\rm e}^{-{\rm i} \l(x)}~, \qquad
\cP ~\to ~ {\rm e}^{{\rm i} \l(x)}\,\cP \, 
{\rm e}^{-{\rm i} \l(x)}~,
\ee
and therefore 
\be
 G (x, x') ~\to ~
{\rm e}^{{\rm i} \l(x)} \,  G (x, x') \,
{\rm e}^{-{\rm i} \l(x')} ~.
\ee

We introduce the proper time representation of $G$:
\be 
G (x, x') ={\rm i} \int_{0}^{\infty} 
{\rm d} s \, K(x,x'|s) ~, 
\ee
where the so-called heat kernel $ K(x,x'|s) $ is formally given by 
\be
 K(x,x'|s) = {\rm e}^{ {\rm i} s \, (\D \,+\,{\rm i} \,\ve)} 
 \, \d^d (x-x') \, {\bf1}~,  \qquad \ve \to + 0~, 
\label{kernel1}
\ee
and which transforms as
\be
K (x, x'|s) ~\to ~
{\rm e}^{{\rm i} \l(x)} \,  K (x, x'|s) \,
{\rm e}^{-{\rm i} \l(x')} 
\label{TL1}
\ee
with respect to the gauge group.

It is advantageous to make use of
the Fourier integral representation
$$
\d^d (x-x') = \int \frac{ {\rm d}^d k }{(2\p)^d}\,
{\rm e}^{{\rm i} \,k. (x-x')}
$$
for the delta-function in (\ref{kernel1}).
To preserve the gauge transformation law (\ref{TL1}), 
however, we should actually represent the full delta-function
$\d^d (x-x') \, {\bf1}$ as follows:
\be
\d^d (x-x') \, {\bf1} = \int \frac{ {\rm d}^d k }{(2\p)^d}\,
{\rm e}^{{\rm i} \,k. (x-x')} \; \cI(x,x')~.
\ee
Here the matrix $\cI(x,x')$ is a functional of the background field 
chosen in such a way that (i) it possesses the gauge transformation law
\be
 \cI (x, x') ~\to ~
{\rm e}^{{\rm i} \l(x)} \,  \cI (x, x') \,
{\rm e}^{-{\rm i} \l(x')} ~;
\label{PDO1}
\ee
(ii) it satisfies the boundary condition 
\be 
\cI (x,x) = {\bf 1}~.
\label{PDO2}
\ee
As a result, the heat kernel takes the form
\be
 K(x,x'|s) =  \int \frac{ {\rm d}^d k }{(2\p)^d}\,
{\rm e}^{{\rm i} \,k. (x-x')} \, 
{\rm e}^{{\rm i} s[ (\nabla +{\rm i}k)^2 + \cP ]} \,
\cI(x,x')
\equiv \hat{K}(x,x'|s) \, \cI(x,x')~.
\label{HK1}
\ee
It is clear that the gauge transformation law 
of the operator $\hat{K}(x,x'|s)$ is
\be
\hat{K} (x, x'|s) ~\to ~
{\rm e}^{{\rm i} \l(x)} \,  \hat{K} (x, x'|s) \,
{\rm e}^{-{\rm i} \l(x)} ~.
\label{TL2}
\ee

Given a {\it gauge invariant} scalar field $\U (x)$ of compact support, 
one can prove the following operator identity:
\be
\hat{K}(x,x'|s) \cdot \hat{\U} (x) = 
\hat{\U} (x') \cdot \hat{K}(x,x'|s)~, \qquad
\hat{\U}(x) \equiv \U(x) \,{\bf 1}~,
\label{master}
\ee
see sect. \ref{super-Green's} for a proof.
This identity implies 
\bea
\hat{K}(x,x'|s) &=& 
\mbox{{\bf :}} \,{\rm e}^{(x'-x). \nabla }\mbox{{\bf :}}
\, \hat{A} (x,x'|s) ~, 
\label{master2}
\eea
where the matrix $\hat{A} (x,x'|s) $ is a functional of the background field
such that 
\be
\hat{A} (x,x'|s) \cdot  \hat{\U} (x) = 
\hat{\U}(x) \cdot \hat{A} (x,x'|s) ~.
\ee
In other words, the matrix $\hat{A} (x,x'|s) $ 
may depend on the covariant derivatives only via multiple 
commutators starting with the master commutators  
$[ \nabla_m, \nabla_n ] = {\rm i}\, F_{mn}$
and $[\nabla_m , \cP ] = (\nabla_m  \cP)$ .
The operator 
$\mbox{{\bf :}} \,{\rm e}^{(x'-x). \nabla }\mbox{{\bf :}}$
is defined by 
\be 
\mbox{{\bf :}} \,{\rm e}^{(x'-x). \nabla }
\mbox{{\bf :}} = 
\sum_{p=0}^{\infty} { 1 \over p!} 
(x'-x)^{m_1} \ldots (x'-x)^{m_p} \, 
\nabla_{m_1} \ldots \nabla_{m_p} 
\label{normal-exp}
\ee
when acting on a gauge covariant field $\vf (x)$ possessing a
covariant Taylor series (see below). More generally, 
the operator  
$\mbox{{\bf :}} \,{\rm e}^{(x'-x). \nabla }\mbox{{\bf :}}$
is defined by 
\be 
\mbox{{\bf :}} \,{\rm e}^{(x'-x). \nabla }\mbox{{\bf :}}\,
\vf (x) = I(x,x') \, \vf (x')~, 
\ee
where $I(x,x')$ is the parallel displacement propagator 
(see below).

It follows from (\ref{master}) that 
in the coincidence limit,
$\hat{K}(x,x|s) $
is not a differential operator, 
\be
\hat{K}(x,x|s) \cdot \hat{\U} (x) = 
\hat{\U} (x) \cdot \hat{K}(x,x|s)~. 
\ee
The latter observation can be restated as the fact that
all covariant derivatives in 
\be 
\hat{K}(x,x|s) =
\int \frac{ {\rm d}^d k }{(2\p)^d}\,
\exp \Big\{ {\rm i} s \, [ (\nabla +{\rm i}k)^2 + \cP ] \Big\} 
\ee
can be organized into multiple commutators upon doing 
the (Gaussian) momentum integration. 

It is well known that the one-loop effective action 
can be expressed via the functional trace of the heat kernel, 
\bea
{\rm Tr} \, K(s) &=& \int {\rm d}^d x \, {\rm tr} \, K(x,x|s)~, \non \\
K(x,x|s) &=& \int \frac{ {\rm d}^d k }{(2\p)^d}\, \Big(
\exp \Big\{ {\rm i} s \, [ (\nabla +{\rm i}k)^2 + \cP ] \Big\} \,
\cI(x,x') \Big)
\Big|_{x'=x} \label{one-loop}
\eea
where  `tr' denotes the trace over gauge group indices. 
In accordance with the above discussion, 
$\hat{K}(x,x|s) $ is not a differential operator, and
we then get
\bea
K(x,x|s) &=& 
\int \frac{ {\rm d}^d k }{(2\p)^d}\,
\exp \Big\{ {\rm i} s \, [ (\nabla +{\rm i}k)^2 + \cP ] \Big\}~, 
\eea
where the boundary condition (\ref{PDO2}) has been taken into 
account. We therefore conclude that, apart from the structural 
requirements (\ref{PDO1}) and (\ref{PDO2}), the concrete choice of 
$\cI(x,x')$ is not relevant at the one-loop level.
 
As already noted, the operator $\hat{A}(x,x'|s)$ depends on 
the gauge covariant derivatives only via commutators. 
In practice, it is quite nontrivial to manifestly organize 
the terms in the expansion of $\hat{A}$ into such commutators. 
It turns out, however, that such an organization occurs
automatically for a special choice of $\cI(x,x')$, 
to be discussed below.
 
Beyond the one-loop approximation, we should work with 
the Feynman propagator $G(x,x')$ at $x\neq x'$. 
Therefore, it would be desirable to choose $\cI(x,x')$ 
in such a way that the heat kernel (\ref{HK1}) takes a 
simple form. The best choice turns out to be the so-called 
parallel displacement propagator $I(x,x')$
along the geodesic connecting 
the points $x'$ and $x$ \cite{DeWitt1}. This object
(which generalizes the Schwinger phase factor \cite{Schw1,Schw2})
satisfies the equation\footnote{The unique solution to the equation 
(\ref{PDO3}) under the boundary condition (\ref{PDO2}) is given by   
$I(x,x') = {\rm P} \, \exp \,(-{\rm i} 
\int_{x'}^{x}  \, A_m \,  {\rm d} x^m )$, 
where the integration is carried out along the straight line 
connecting the points $x'$ and $x$. This 
explicit form of $I(x,x')$ is not required for actual loop 
calculations.} 
\be
(x-x')^a \, \nabla_a \, I(x,x') = 0~,
\label{PDO3}
\ee
which implies
\be 
(x'-x)^{a_1} \ldots (x'-x)^{a_n} \, 
\nabla_{a_1} \ldots \nabla_{a_n} \, I(x,x') = 0~, 
\label{PDO4}
\ee
for any positive integer $n$. 
The latter means that  the operator 
$\mbox{{\bf :}} \,{\rm e}^{(x'-x). \nabla }\mbox{{\bf :}}$
in (\ref{master2})
collapses to the unit operator when acting on $I(x,x')$. 
It is worth noting that the identities (\ref{PDO4}) lead to 
\be 
\nabla_{(a_1} \ldots \nabla_{a_n)} \, I(x,x') \Big|_{x=x'} =0~,
\label{PDO5}
\ee
for any positive integer $n$.

In what follows, we identify $\cI(x,x')$ with the 
parallel displacement propagator\footnote{Representation 
(\ref{HK1}) with $\cI(x,x')$ chosen to be 
the parallel displacement propagator, was the starting 
point of Avramidi's analysis \cite{Avr1}.}. 
${}$From the equation (\ref{PDO3}), 
the transformation law (\ref{PDO1}) and the boundary 
condition (\ref{PDO2}), one can easily deduce \cite{BV}
\be 
I(x,x') \, I(x',x) = {\bf 1}~.
\ee
Under Hermitian conjugation, this object transforms as
\be 
\Big( I (x,x') \Big)^\dagger = I(x' , x)~.
\ee

Let $\vf (x)$ be a field transforming in some representation 
of the gauge group. It can be expanded in a 
{\it covariant Taylor series} \cite{BV}
\be 
\vf (x) = I(x,x') \, \sum_{n=0}^{\infty} 
{1 \over n!} \, (x-x')^{a_1} \ldots (x-x')^{a_n} \, 
\nabla _{a_1} \ldots \nabla_{a_n} \vf (y) \Big|_{y = x'}~,
\label{CTS1}
\ee  
see sect. \ref{super-parallel-dis} for a proof.
We can apply this expansion to $\nabla_b I(x,x')$ 
considered as a function of $x$: 
\be 
\nabla_b I(x,x') = I(x,x') \, \sum_{n=0}^{\infty} 
{1 \over n!} \, (x-x')^{a_1} \ldots (x-x')^{a_n} \, 
\nabla _{ a_1} \ldots \nabla_{a_n} \nabla_b 
I (y, x') \Big|_{y = x'}~.
\ee
Using the identity
$$
\nabla _{( a_1} \ldots \nabla_{a_n} \nabla_{b)} 
I (y, x') \Big|_{y = x'} = 0
$$
in conjunction with the commutation relation
$[ \nabla_m, \nabla_n ] = {\rm i}\, F_{mn}$, 
one can express 
$$\nabla _{( a_1} \ldots \nabla_{a_n )} \nabla_b 
I (y, x') \Big|_{y = x'}
$$ 
in terms of covariant derivatives
of the field strength (see the Appendix for more
detail).  
One ends up with \cite{Avr1}
\bea 
\nabla_b I(x,x') &=& {\rm i} \, I(x,x') \, \sum_{n=1}^{\infty} 
{ n  \over (n+1)!} \, 
(x-x')^{a_1} 
\ldots 
(x-x')^{a_{n-1}}
(x-x')^{a_n}  
\non \\ 
&\times &
\nabla'_{a_1} \ldots \nabla'_{a_{n-1} } F_{a_n \,b } (x')~.
\label{PTO-der1}
\eea
This relation has a fundamental significance
in our considerations. 
It expresses covariant derivatives of 
$I(x,x') $ in terms of $I(x,x')$ itself and covariant derivatives of 
the field strength $F_{ab}$. 
This is the property which automatically results in 
the organization of the covariant derivatives in $\hat{A}(x,x'|s)$
into commutators. 

Relation (\ref{PTO-der1}) can be rewritten in an integral form 
\bea
\nabla_b I(x,x') &=& {\rm i} \, \int_{0}^{1} {\rm d} t\,
I(x,x(t)) \, \frac{\pa x^d (t) }{\pa x^b} \, \dot{x}^c (t)\,
F_{cd}(x(t)) \, 
I(x(t), x') ~, \label{integralform} \\
x(t) &=& (x-x')\,t + x'~, \non 
\eea
which was derived, in a more general setting, 
many years ago in \cite{LM}. Applying the covariant Taylor 
expansion to $I(x,x(t)) \, F_{cd}(x(t)) $, 
\bea 
&&\qquad \qquad  I(x,x(t)) \,F_{cd}(x(t)) \, I(x(t),x) \non \\
& = &\sum_{n=0}^{\infty} 
{(-1)^n \over n!} \, 
(1-t)^n \,(x-x')^{a_1} \ldots (x-x')^{a_n} \, 
\nabla _{a_1} \ldots \nabla_{a_n} F_{cd}(x)~, 
\eea
and making use of the identity
\be 
I(x,x(t)) \, I(x(t),x') = I(x,x')~, 
\ee
eq. (\ref{integralform}) leads to 
\bea 
\nabla_b I(x,x') &=& - {\rm i} \, \sum_{n=1}^{\infty} 
{ (-1)^n  \over (n+1)!} \, 
(x-x')^{a_1} 
\ldots 
(x-x')^{a_{n-1}}
(x-x')^{a_n}  
\non \\ 
&\times &
\nabla_{a_1} \ldots \nabla_{a_{n-1} } F_{a_n \,b } (x)
\, I(x,x')~.
\label{PTO-der-new}
\eea
Of course, the latter relation can be deduced directly 
from (\ref{PTO-der1}).
    
It is worth noting that the relation (\ref{PTO-der1}) and the delta 
function representation 
\be
\d^d (x-x') \, {\bf1} = \int \frac{ {\rm d}^d k }{(2\p)^d}\,
{\rm e}^{{\rm i} \,k. (x-x')} \; I(x,x')
\ee
make quite obvious 
the standard property of the delta-function,
\be
\nabla_a \Big( \d^d (x-x') \, \d^i{}_{i'} \Big) 
= - \nabla'_a \Big( \d^d (x-x') \, \d^i{}_{i'} \Big) ~.
\ee

Relations (\ref{PTO-der1}) and (\ref{PTO-der-new})
simplify drastically in the case 
of a covariantly constant gauge field, 
\be 
\nabla_a F_{bc} = 0~.
\ee
Then, eqs. (\ref{PTO-der1}) and (\ref{PTO-der-new}) take the form
\be
\nabla_b I(x,x') = { {\rm i} \over 2} \, (x-x')^a \,
I(x,x') \,
F_{ab}(x') 
= { {\rm i} \over 2} \, (x-x')^a \,F_{ab}(x) \,I(x,x')~.
\ee

Let us fix some space-time point $x'$ and consider the 
following gauge transformation: 
\be
{\rm e}^{{\rm i} \l(x)}  = I (x', x) ~, \qquad 
{\rm e}^{{\rm i} \l(x')} =  {\bf 1}~. 
\ee
Applying this gauge transformation to $I(x,x')$, 
as in eq. (\ref{PDO1}), the result is
\be
I(x,x') = {\bf 1}~, 
\label{F-S1}
\ee
which is equivalent, due to (\ref{PDO3}), 
to the Fock-Schwinger gauge \cite{Fock,Schw1} 
\be
(x-x')^m \,  A_m(x) = 0~.
\label{F-S2}
\ee
In the Fock-Schwinger gauge, the relation (\ref{PTO-der1})
becomes \cite{Shif}
\be
A_b (x)  =  \sum_{n=1}^{\infty} 
{ n  \over (n+1)!} \, 
(x-x')^{a_1} 
\ldots 
(x-x')^{a_{n-1}}
(x-x')^{a_n}  
\nabla'_{a_1} \ldots \nabla'_{a_{n-1} } F_{a_n \,b } (x')~.
\ee
Thus, all coefficients in the Taylor expansion of $A(x)$  acquire a 
geometric meaning.

We now comment on the explicit evaluation of the kernel
\be
K(x,x'|s) =  \int \frac{ {\rm d}^d k }{(2\p)^d}\,
{\rm e}^{{\rm i} \,k. (x-x')} \,
{\rm e}^{{\rm i} s[ (\nabla +{\rm i}k)^2 + \cP ]} \,
I(x,x')~.
\label{starting}
\ee
It should be pointed out that we are interested in  
a manifestly covariant expansion of the heat kernel which can be truncated 
at required order in the derivative expansion.
Rescaling $k$, the right hand side of (\ref{starting}) is
the result of applying  the operator 
\be
\frac{1}{(4 \p^2 s)^{d/2}} \,\int {\rm d}^d k \,
{\rm e}^{- {\rm i} k^2 + {\rm i}\, s^{-1/2} \,k. (x-x')} \,
{\rm e}^{[{\rm i} s \nabla^2  - 2 s^{1/2} k. \nabla + {\rm i} \,s
\cP ]}
\label{starting2}
\ee
to the parallel displacement propagator. The second exponential factor 
here should  then be expanded in a Taylor series. 
Whenever  a covariant derivative $\nabla_b$ 
 from this series
hits $I(x,x')$, we  apply 
the relation  (\ref{PTO-der1}). Given a product of 
the form $\cP (x) \,I(x,x')$, we represent it as 
\be
\cP (x) \,I(x,x') = 
I(x,x') \, \sum_{n=0}^{\infty} 
{1 \over n!} \, (x-x')^{a_1} \ldots (x-x')^{a_n} \, 
\nabla'_{a_1} \ldots \nabla'_{a_n} \cP (x') ~.
\label{P}
\ee
A generic term in
the Taylor expansion will involve a Gaussian moment of the form
\be
\langle k^{a_1} \ldots  k^{a_n} \rangle \equiv
\frac{1}{(4 \p^2 s)^{d/2}} \,\int {\rm d}^d k \,
{\rm e}^{- {\rm i} k^2 + {\rm i}\, s^{-1/2} \,k. (x-x')} \,
s^{1/2} k^{a_1} \ldots s^{1/2} k^{a_n}~,
\label{moment}
\ee
where each $k^{a_i}$ comes together with an 
$s$-independent factor of $\nabla_{a_i}$;
there also occur insertions of $s \nabla^2$ and $s \cP$.
To compute the moments (\ref{moment}), we can introduce a generating 
function $Z(J)$, 
\bea
Z(J) &=& \frac{1}{(4 \p^2 s)^{d/2}} \,\int {\rm d}^d k \,
{\rm e}^{- {\rm i} k^2 + {\rm i}\, s^{-1/2} \,k. (x-x') \, +
s^{1/2}\, J.k}~, \non \\ 
\langle k^{a_1} \ldots  k^{a_n} \rangle &=&
\frac{\pa^n}{\pa J_{a_1} \ldots \pa J_{a_n} }\, Z(J)\Big|_{J=0}~.
\eea
One readily gets 
\be
Z(J) = \frac{\rm i}{( 4 \p {\rm i}
s)^{d/2}} \, {\rm e}^{{\rm i} (x-x')^2/4s} \, {\rm e}^{- {\rm i} s
J^2/4 + J.(x-x')/2}~.
\label{genf}
\ee
Then, the resulting expression for the kernel is of the form
\be
K(x,x'|s) =
\frac{\rm i}{( 4 \p {\rm i} s)^{d/2}} \, 
{\rm e}^{{\rm i} (x-x')^2/4s}\, F(x,x'|s)~, \qquad 
 F(x,x'|s) = \sum_{n=0}^{\infty} a_n (x,x') \,({\rm i}s)^n~,
\ee
where\footnote{Had one started with the more general representation
(\ref{HK1}) instead of (\ref{starting}), the result for $a_0$
would then have been
$a_0 (x,x') =
\mbox{{\bf :}} \,{\rm e}^{(x'-x). \nabla }\mbox{{\bf :}}
\, \cI(x,x') = I (x,x')$.}
\be 
a_0 (x,x') =
\mbox{{\bf :}} \,{\rm e}^{(x'-x). \nabla }\mbox{{\bf :}}
\, I(x,x') = I (x,x')~. 
\ee
This is consistent with the
standard Schwinger-DeWitt asymptotic expansion of the 
heat kernel \cite{DeWitt1}. By construction, the coefficients
$a_n$ have the form 
\bea
a_n(x,x') &=& {\bf a}_n\Big( F(x), \nabla F (x), \ldots , 
\cP(x) , \nabla \cP (x) \ldots ; x-x' \Big) \, I(x,x') \non \\
&=& I(x,x')\,                       
{\bf a}'_n \Big( F(x'), \nabla' F (x'), \ldots , 
\cP(x') , \nabla' \cP (x') \ldots ; x-x' \Big) ~,
\eea
where the functions ${\bf a}_n$ and ${\bf a}'_n$ are straightforward
to compute using the scheme described above. 
In the standard Schwinger-DeWitt technique, 
one has to solve the recurrence relations which follow 
from the equation 
\be 
\Big( {\rm i} \,{\pa \over \pa s}  +\nabla^2 + \cP \Big) 
K(x,x'|s) = 0~.
\ee
In practice, these recurrence relations prove more difficult to solve
than to implement the above perturbation scheme.

\sect{Evaluating multi-loop diagrams}

In sect. 2, a covariant procedure for the derivative expansion of Green's 
functions  in the background field method 
was presented. We now extend this to the evaluation of Feynman diagrams. 

As already discussed in sect. 2, at the one loop level, 
the effective action can be expressed in terms 
of the functional trace of the heat kernel, and therefore only a 
knowledge of the coincidence limit of the kernel is required. It was 
noted that in principle this does not require a 
specific choice of  $\cI(x,x')$ in (\ref{one-loop})
because, after doing the momentum integration, all covariant 
derivatives assemble into commutators and there are actually no derivatives of 
$\cI(x,x')$ to be evaluated. In practice, it is quite 
difficult to assemble the commutators, and it proves  efficient to 
use the procedure outlined in sect. 2: namely,
to choose $\cI(x,x')$ to be the parallel displacement propagator 
$I(x,x')$ and to evaluate the action of covariant derivatives  
on $I(x,x')$ using either (\ref{PTO-der1})
or (\ref{PTO-der-new}). This automatically generates 
the required commutators.

Beyond one loop, the situation is more complicated, in that  
diagrams involve multiple Green's functions, and interaction vertices
will in general include  insertions of background fields and covariant 
derivatives. When covariant derivatives act on Green's functions,
the generic structure will be of the form 
\bea
&&  \nabla_{a_1} \cdots \nabla_{a_{n}} 
 \nabla'_{b_1} \cdots \nabla'_{b_{m}}
  K(x,x'|s)  = 
 \int \frac{ {\rm d}^d k }{(2\p)^d}\,
{\rm e}^{{\rm i} \,k. (x-x')} \,(\nabla + {\rm i} k)_{a_{1}} \cdots 
(\nabla + {\rm i} k)_{a{_n}} 
\nonumber \\
& & \qquad \qquad \times  
{\rm e}^{{\rm i} s[ (\nabla +{\rm i}k)^2 + \cP ]} \,
(\nabla' - {\rm i} k)_{b_1} \cdots (\nabla' - {\rm i} k)_{b_{m}} \,
I(x,x')~, 
\label{dKernel}
\eea
where the proper time integral has been omitted.
When $\nabla$ hits $I(x,x')$, the result can be represented 
as either (\ref{PTO-der1}) or (\ref{PTO-der-new}).  
When $\nabla'$ hits $I(x,x')$, the result can be 
evaluated using 
\be 
\nabla'_b \, I(x,x') = - I(x,x') \, \nabla'_b I(x',x) \, I(x,x')~.
\ee 
After having applied all the covariant derivatives
to $I(x,x')$, the integrand in (\ref {dKernel}) 
can be expressed either as
\be 
{\rm e}^{ {\rm i} \,[ k. (x-x') -s\,k^2 ]}\, 
\J \Big( F(x) , \nabla F (x), \ldots, \cP (x) , 
\nabla  \cP (x)  , \ldots ; x-x' , k, s\Big) \, I(x,x')
\label{dKernel2}
\ee
or 
\be 
{\rm e}^{ {\rm i} \,[ k. (x-x') -s\,k^2 ]}\, 
I(x,x') \, \J' \Big( F(x') , \nabla' F (x') , \ldots,
\cP(x') ,\nabla'  \cP (x'), \ldots ; x-x' , k, s\Big) ~.
\label{dKernel3}
\ee

At this stage, there are two possible ways to proceed. 
${}$For any Feynman graph, 
the momentum integrals 
in {\it all} kernels (\ref {dKernel})  can either be
(i) left to the end of calculation; or 
(ii) carried out first. 
These choices give rise to two different {\it manifestly} covariant
perturbation schemes: (i) momentum space scheme;
(ii) configuration space scheme. In our opinion, 
the former is most suitable for arbitrary backgrounds, 
while the latter is better adapted to special background
configurations such as a covariantly constant field. 
We will describe in detail the momentum space scheme. 
It is worth noting that the configuration space scheme
was advocated in \cite{BV,BV2,BvdV}.

${}$For the purposes of illustration, we first consider two-loop 
contributions to the effective action. 
The relevant Feynman diagrams are either of  the `eight' 
type  or `fish'  type, as shown in Figure 1. 
\begin{figure}[!htb]
\begin{center}
\includegraphics{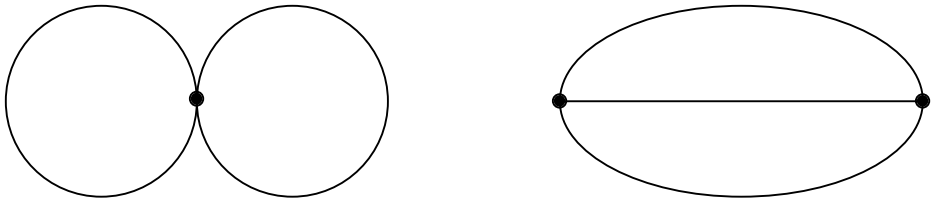}
\caption{Two-loop graphs: `eight' diagram and `fish' diagram.}
\end{center}
\end{figure}

The `eight' diagram 
is relatively simple to treat since 
it involves the product of two Green's functions 
or their covariant derivatives in a
coincidence limit.
The situation for the `fish' diagram is somewhat more complex.
This diagram  makes a contribution to the effective action of the 
form (with gauge indices suppressed) 
\be
\int\limits_0^{\infty} {\rm d} s_1  
\int\limits_0^{\infty} {\rm d} s_2 
 \int\limits_0^{\infty} {\rm d} s_3 
\int {\rm d}^d x \, \int {\rm d}^d x' \,  V_1(x) \, V_2(x') \, 
K_1(x,x'|s_1) \,K_2(x,x'|s_2) \, K_3(x,x'|s_3)~,
\label{fish}
\ee
where the vertex factors $V_1(x)$ and $V_2(x')$  will in general 
contribute insertions of the background fields, and may also contain 
covariant derivatives which act on one or more of the kernels. 
The procedure for evaluating (\ref{fish}), to be outlined below, 
is to reduce this diagram to 
a `skeleton'. The latter is characterized 
by the following properties: 
(i) all explicit background field dependence is collected 
at a single vertex, say $x'$; 
(ii) the  vertex at $x$ is connected to that
at $x'$ by four (undifferentiated) parallel 
displacement propagators only; 
(iii) the skeleton is gauge invariant at each vertex. 

The kernels and their derivatives in (\ref{fish}) 
should be expressed as in (\ref{dKernel})  and (\ref{dKernel3}). 
In particular, the background field dependence in the 
kernels coming from either $\cP$ or from covariant derivatives 
of $I(x,x')$  are
evaluated at $x'$, with $\cP (x) \,I(x,x')$ treated as in 
(\ref{P}) .
The  $x$-dependence of each of the  kernels is then only via a 
multiplicative factor $I(x,x')$, 
Taylor factors $(x-x')^n$,  
and via a phase factor ${\rm e}^{ {\rm i} k_i.(x-x')}$, where $k_i$  
$(i = 1,2\, {\rm or} \, 3)$  denotes 
the momentum associated with the  kernel. 
In contrast, the vertices $V_1(x)$ and $V_2(x')$ 
contribute background fields 
evaluated at  $x$ and $x'$ respectively. As stated earlier, the aim is to get 
all explicit dependence on the background to be in the form of fields 
evaluated at $x'.$ This can be achieved by the covariant Taylor 
expansion of the product of fields in $V_1(x)$ 
in terms of fields evaluated at $x'$, 
\be 
\cV_1 (x) = I(x,x') \, \sum_{n=0}^{\infty} 
{1 \over n!} \, (x-x')^{a_1} \ldots (x-x')^{a_n} \, 
\nabla' _{a_1} \ldots \nabla'_{a_n} \cV_1 (x') ~,
\ee  
where $\cV_1(x)$ denotes the result of removing all covariant 
derivatives from the vertex $V_1(x)$.
The covariant Taylor  expansion introduces 
an additional factor of the 
parallel displacement  propagator $I(x,x')$ 
(in the representation of the gauge 
group under which  $\cV_1$ transforms). 
This is denoted by the dashed 
line $I_4(x,x')$ in Figure 2, while the 
solid lines $I_i(x,x'), i = 1,2,3$ denote 
the three parallel displacement propagators 
arising from the kernels. The filled square denotes the 
dependence on background fields, evaluated only at $x'.$
\begin{figure}[!htb]
\begin{center}
\includegraphics{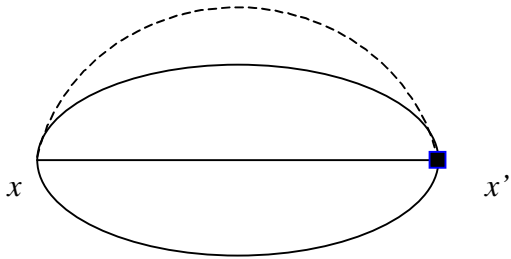}
\caption{Accumulation of explicit background field dependence at 
one point.}
\end{center}
\end{figure}

The parallel displacement propagators $I_1(x,x'), \cdots, I_4(x,x')$ 
in Figure 2 are in general in different representations of the gauge 
group. However, their  product 
\be
I_4(x,x') \, I_1(x,x') \, I_2(x,x') \, I_3(x,x')
\ee
(gauge indices suppressed) is gauge invariant at  $x.$  
Since each of these propagators satisfies the same 
equation (\ref{PDO3}), it is straightforward to show
\be
I_4(x,x') \, I_1(x,x') \, I_2(x,x') \, I_3(x,x')
= 1~.
\ee
More precisely, the right-hand side of this relation is a product of 
Kronecker deltas.
 Thus the `skeleton' of parallel displacement 
propagators collapses, leaving all background field dependence in the 
form of fields evaluated at $x',$ depicted by the square in Figure 2.
There are also explicit factors of $(x-x')$ 
from the covariant Taylor expansion of fields in 
the vertex at $x$, and from covariant derivatives of the 
parallel displacement propagators. 
These can be traded for momentum derivatives by using 
\be 
(x-x')^a \, {\rm e}^{ {\rm i} k.(x-x')} F(k)  =  - \, {\rm i} \, F(k) \,
\frac{\partial}{\partial k_a} \, {\rm e}^{ {\rm i} k.(x-x')} 
={\rm i}  \, {\rm e}^{ {\rm i} k.(x-x')} \, \frac{\partial}{\partial 
k_a} F(k) +\ldots ~,
\label{kderiv} 
\ee
where the dots denote
a total derivative, 
which does not contribute to the momentum integral. This  leaves all the 
dependence on $(x-x')$ in the form of the phase factor
$$
{\rm e}^{{\rm i} \, (k_1 + k_2 + k_3).(x-x')}~.
$$
 Since all background fields now depend 
only on $x',$ the integral over $x$ is
\be
\int {\rm d}^dx \,{\rm e}^{ {\rm i} (k_1 + k_2 + k_3).(x - x')} =
 (2 \p)^d \, \delta^d(k_1 + k_2 + k_3)~,
 \ee
and this delta function allows  one of the three Gaussian momentum 
integrals associated with the kernels to be eliminated. 
The procedure outlined above results in an expression for the 
contribution to the effective action from  the `fish' 
diagram in the form\footnote{All covariant regularization 
schemes are consistent with our approach, therefore this issue
does not require elaboration.} 
\bea 
& & \int {\rm d}^d x' 
\int\limits_0^{\infty} {\rm d} s_1 
\int\limits_0^{\infty} {\rm d} s_2 
\, \int\limits_0^{\infty} {\rm d} s_3 
\int \frac{{\rm d}^dk_1}{(2 \p)^d} 
\, \frac{{\rm d}^dk_2}{(2 \p)^d} \, 
{\rm e}^{- {\rm i}\, s_1 k_1^2} \,
{\rm e}^{- {\rm i}\, s_2 k_2^2}
 \,{\rm e}^{- {\rm i}\, s_3 (k_1 + k_2)^2}
\nonumber \\
& & 
\phantom{ \int {\rm d}^d x' 
\int\limits_0^{\infty} {\rm d} s_1 
\int\limits_0^{\infty} {\rm d} s_2 
\, \int\limits_0^{\infty} {\rm d} s_3 } 
\times  F(s_1, s_2, s_3, k_1, 
 k_2, \phi(x'))~,
 \eea
where $\phi$ denotes background fields and 
their covariant derivatives.
The momentum integrals here are just Gaussian moments, 
which are straightforward to compute.

At three and higher loops, a similar procedure can be implemented, 
although the presence of extra vertices adds a complication. The aim 
is to transform the $n$-vertex diagram to a `skeleton' which 
has the following properties:
(i) all explicit background field dependence is collected 
at a single vertex, say $x_n$; 
(ii) the remaining vertices at $x_1, \ldots , x_{n-1}$ are connected to each 
other and to the vertex at $x_n$ by (undifferentiated) parallel 
displacement propagators only; 
(iii) the skeleton is gauge invariant at each vertex. 
Consider, for example, the three-loop graph  shown in 
Figure 3.
\begin{figure}[!ht]
\begin{center}
\includegraphics{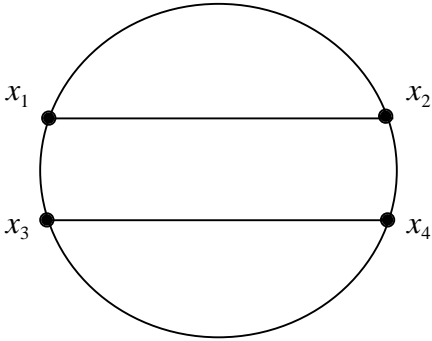}
\caption{A three-loop graph.}
\end{center}
\end{figure}

\noindent
Suppose we wish to move all the background field dependence to 
a single vertex, say  at $x_4.$
As a first step, we could choose to use the covariant Taylor expansion to 
express the background fields in the vertex $V_1(x_1)$ in terms of fields 
evaluated at $x_2,$ resulting in the introduction of a parallel 
displacement propagator $I(x_1,x_2)$ in the representation of the 
gauge group under which the fields in $V_1$ transform. Using the 
steps outlined for the `fish' diagram, it is  also possible to 
manipulate the kernel $K(x_1, x_2|s_1)$ (or its derivatives if 
 derivatives from the vertices   act on it) and the 
explicit factors of $(x_1 - x_2)$ in the Taylor series
 so that all of the $x_1$ 
dependence is in the form $e^{{\rm i} k_1.(x_1 - x_2)} \, I(x_1, 
x_2).$ Here, the factor of $I(x_1, x_2)$ is in the representation under 
which $K(x_1, x_2|s_1)$ transforms. A new vertex, $\tilde{V}_2(x_2),$ 
has effectively been generated at $x_2,$ consisting of the original 
background dependence of the vertex multiplied by the additional 
background dependence generated from $V_1(x_1)$ and $K(x_1, x_2|s_1).$
Diagrammatically, we have the situation in Figure 4, where the 
circles denote original vertices and the square denotes the new 
effective vertex.
\begin{figure}[!ht]
\begin{center}
\includegraphics{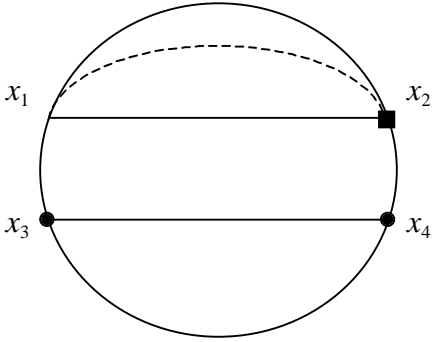}
\caption{Removing explicit field dependence at $x_1$.}
\end{center}
\end{figure}
 
Repeating this procedure, the explicit dependence on background 
fields can be progessively moved through the diagram, at each step 
removing the field dependence at one vertex at the cost of 
introducing an additional parallel displacement propagator.
It should be stressed that this procedure preserves manifest gauge 
invariance at each vertex.
When all explicit background field dependence has been moved to a 
single point, the result is a skeleton as defined above. 

Having produced the skeleton from 
the original Feynman diagram, it is now necessary to evaluate it. 
A very special case is that of a two-vertex skeleton,
as illustrated in Figure 
5. 
\begin{figure}[!ht]
\begin{center}
\includegraphics{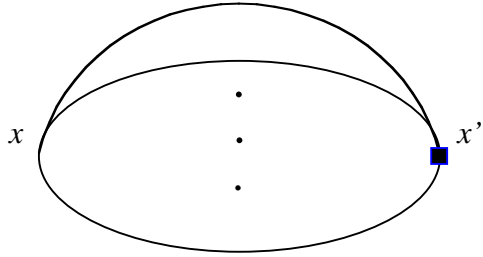}
\caption{An arbitary two-vertex skeleton.}
\end{center}
\end{figure}
Here, the product of the parallel diplacement propagators  
is gauge invariant at $x.$ 
Since  all the propagators  satisfy the same 
equation (\ref{PDO3}), it is straightforward to show that the product  
of parallel diplacement propagators simply collapses to 
a product of Kronecker deltas, 
leaving a gauge invariant contribution at a single point, $x'$.
In the case where the skeleton contains three or more vertices, it is 
possible to systematically reduce it to a skeleton with only two 
vertices; the resulting diagram then collapses to a single point as 
above. We will 
explain the procedure for reducing a skeleton with $n$ vertices to 
one with $n-1$ vertices; this process is then iterated to produce the 
two vertex skeleton.

The generic structure of the skeleton with $n$ vertices is shown in 
Figure 6, where $x_1$ is the point where
all the explicit background field dependence is accumulated, 
and the lines are parallel displacement 
propagators. 
\begin{figure}[!htb]
\begin{center}
\includegraphics{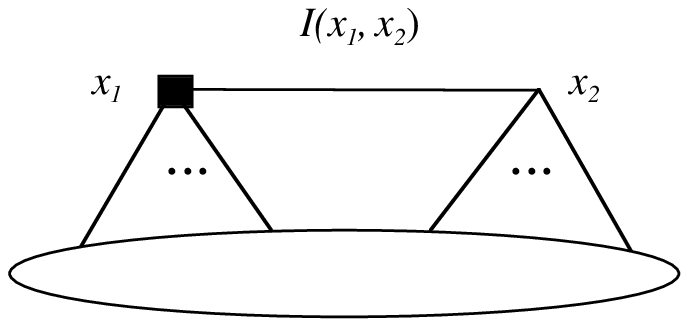}
\caption{An arbitrary $n$-vertex skeleton, $n \geq 3.$}
\end{center}
\end{figure}
Focusing attention on a vertex adjacent to that at $x_1,$ 
say that at  $x_2,$ the diagram has  the structure
\be
{\rm tr} \Big( I(x_1, x_2) \, \chi(x_2, x_1) \Big),
\ee
where $\chi(x_2,x_1)$ consists of the vertex at $x_1$ and products of parallel 
displacement propagators (dependence on points other than $x_1$ 
and $x_2$ has been suppressed).
We can make use of the covariant 
Taylor series to expand $I(x_1, x_2) \, \chi(x_2, x_1)$ in the form
\be 
I(x_1, x_2) \, \chi(x_2, x_1) = \sum_{n=0}^{\infty} \frac{1}{n!} \, 
(x_2 - x_1)^{a_1} \cdots (x_2 - x_1)^{a_n} \, 
\nabla _{a_1} \ldots \nabla_{a_n} \chi(y, x_1) \Big|_{y = x_1}~.
\ee
By this means, all $x_2$ dependence of the skeleton is now in the form of factors $(x_2 - x_1).$ 
These factors can be replaced by momentum derivatives of the phase factor 
${\rm e}^{{\rm i} k. (x_1 - x_2)}$ associated with a Green's function
 in the original Feynman diagram, as in (\ref{kderiv}). 
This process reduces number of vertices in the skeleton by one, as 
illustrated in Figure 7.
\begin{figure}[!htb]
\begin{center}
\includegraphics{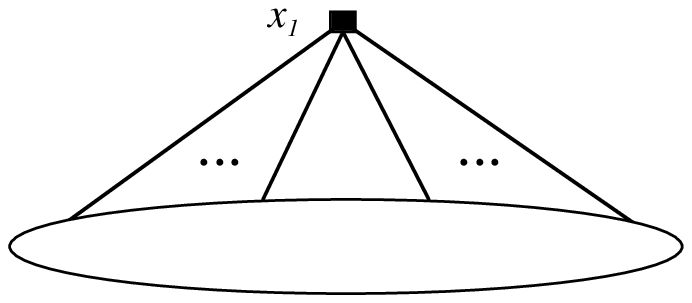}
\caption{Reducing an $n$-vertex skeleton to an $(n-1)$-vertex 
skeleton, $n \geq 3.$}
\end{center}
\end{figure}
This completes our prescription for computing multi-loop 
contributions to the effective action
in a manifestly covariant manner.

In the literature \cite{BV,BV2}, the configuration
space approach has been the favoured means to achieve a covariant
derivative expansion of the effective action.
A possible explanation for this is a belief that the transition 
to momentum space should be incompatible with manifest 
gauge invariance, see e.g. \cite{BvdV}. 
As we have shown above, the latter is not the case.
Our momentum space scheme
is well-adapted to the derivative
expansion of the effective action, 
in that it automatically links the 
factors of momentum  to covariant derivatives of the
background fields. Further, this approach is more economical in that 
there is no need to do
separate configuration and momentum space integrals
(configuration space integrals are just trivial).
In contrast, the configuration
space approach
requires separate computation (by carrying out momentum integrals) 
of each Greeen's function
in a diagram to the required order in the derivative expansion, and
then the sewing together of
the Green's functions with configuration space
integrals. 
As noted earlier, the configuration space approach 
is well suited to the computation of the effective action 
in special backgrounds,  as an exact
expression for the Green's function can sometimes be computed. An 
example is a covariantly constant background; 
we treat this case in $\cN =1$ super Yang-Mills in
sect. 6. It is worth noting that the momentum space scheme
can easily be formulated in curved space due to the existence 
of a covariant Fourier integral \cite{Avr1}.

\sect{Parallel displacement  propagator in superspace}
\label{super-parallel-dis}

In four space-time dimensions, $d=4$, we consider a flat global
$\cN$-extended superspace parametrized by variables 
$z^m = (x^m, \q^\m_i , {\bar \q}_{\dot{\m}}^i)$, where $i= 1,\ldots, \cN$.
We recall that 
the flat covariant derivatives $D_A = (\pa_a, D_\a^i , {\bar D}^\ad_i)$
are related to the supersymmetric Cartan 1-forms 
$\o^A =  (\o^a , \o^\a_i , {\bar \o}_\ad^i ) $ by 
\be
{\rm d}z^M \,\pa_M = \o^A \, D_A~, \qquad
\o^A 
= ({\rm d} x^a - {\rm i} \, {\rm d}\q_i  \s^a   {\bar \q}^i
+  {\rm i} \, \q_i  \s^a  {\rm d} {\bar \q}^i , 
{\rm d} \q^\a_i , {\rm d} {\bar \q}_\ad^i )~, 
\ee
and satisfy  the algebra
\be 
[D_A, D_B \} = T_{AB}{}^C \, D_C~, 
\ee
where the only non-vanishing components of the {\it constant}
torsion $T_{AB}{}^C $ correspond to the anticommutator 
$\{ D_\a^i , {\bar D}_{\bd j} \} 
= -2 {\rm i}\, \d^i{}_j \, (\s^c)_{\a \bd}  \, \pa_c$.

To describe a Yang-Mills supermultiplet, 
we introduce gauge covariant derivatives 
$\cD_A = ( \cD_a, \cD_\a^i, {\bar \cD}^\ad_i )$, 
\be
\cD_A = D_A +{\rm i}\, \G_A(z) ~, \qquad
[\cD_A, \cD_B \} = T_{AB}{}^C \, \cD_A + 
{\rm i}\,F_{AB}(z) ~, 
\label{CDA}
\ee
where the connection $\G_A$ takes its values in the Lie algebra 
of a compact gauge group. The operators $\cD_A$ possess 
the following gauge transformation law
\be
\cD_A ~\to ~ {\rm e}^{{\rm i} \t(z)} \, \cD_A\,
{\rm e}^{-{\rm i} \t(z)}~, \qquad 
\t^\dagger = \t ~, 
\label{tau}
\ee
with the gauge parameter $\t(z)$ being arbitrary modulo 
the reality condition imposed. 

In the case of $\cN=1$ supersymmetry, it is known that 
one can always bring, by applying a complex gauge transformation, 
the covariant derivatives 
$\cD_A = (\cD_a, \cD_\a, {\bar \cD}^\ad) $ to 
the so-called covariantly chiral representation where 
\be 
\cD_\a = {\rm e}^{-V(z)}\, D_\a \,{\rm e}^{V(z)}~, \qquad  
{\bar \cD}_\ad = {\bar D}_\ad ~, \qquad V^\dagger = V~, 
\ee
with $V$ the gauge prepotential.
In the chiral representation (sometimes called the $\L$-frame), 
the $\t$-transformations
(\ref{tau}) are replaced by chiral $\L$-transformations
defined by 
\be
\cD_A ~\to ~ {\rm e}^{{\rm i} \L(z)} \, \cD_A\,
{\rm e}^{-{\rm i} \L(z)}~, \qquad 
{\bar D}_\ad \L =0 ~.  
\label{lambda}
\ee
In the present paper, we always work, for definiteness, in the $\t$-frame. 
It is also worth noting that one can define
an analogue of the $\L$-frame in the case of $\cN=2$ supersymmetry
at the cost of embedding the standard $\cN=2$ superspace 
into the so-called harmonic superpace \cite{GIOS}.

Let $z^M (t) = (z-z')^M \, t+z'^M$ be the straight line connecting 
two points $z$ and $z'$ in superspace, parametrized 
such that $z^M (0) = z'^M$ and 
$z^M (1) = z^M $. We then have $\dot{z}^M \, \pa_M =\z^A \, D_A$, where
the two-point function 
$\z^A \equiv \z^A (z,z') = -\z^A (z',z) $ is
\bea
 \z^A = 
\left\{
\begin{array}{l}
\z^a = (x-x')^a - {\rm i} (\q-\q')_i \s^a {\bar \q}'^i 
+ {\rm i} \q'_i \s^a ( {\bar \q} - {\bar \q}')^i ~, \\
\z^\a_i = (\q - \q')^\a_i ~, \\
{\bar \z}_\ad^i =({\bar \q} -{\bar \q}' )_\ad^i ~. 
\end{array} 
\right. 
\label{super-interval}
\eea
Given a gauge invariant superfield $U(z)$, for
$U(t) = U(z(t))$ we have 
\be
\frac{{\rm d}^n U}{{\rm d} t^n} = 
\z^{A_n} \ldots \z^{A_1} \, 
D_{A_1} \ldots D_{A_n} \, U~. 
\ee
This leads to a {\it supersymmetric Taylor series}
\be
U(z) = \sum_{n=0}^{\infty} {1 \over n!}\,
\z^{A_n} \ldots \z^{A_1} \, 
D'_{A_1} \ldots D'_{A_n} \, U(z') ~. 
\label{super-Taylor1}
\ee

By definition, the parallel displacement propagator along 
the straight line, $I(z,z')$, is specified by the 
requirements:\footnote{The unique solution to the requirements
(i)--(iii) is given by   
$I(z,z') = {\rm P} \, \exp \,[-{\rm i} 
\int_{z'}^{z} {\rm d}t \, \z^A \G_A (z(t)) ]$, 
where the integration is carried out along the straight line 
connecting the points $z'$ and $z$. For an arbitrary path
$\g: t \to z^M(t) $ in superspace, such
supersymmetric phase factors of the form
$ \S[\g] = {\rm P} \, \exp \,[-{\rm i} 
\int_{\g} \o \cdot \G (z(t)) ]$, were considered 
in \cite{Gates,GJN,MM}.}  \\
(i) the gauge transformation law
\be
 I (z, z') ~\to ~
{\rm e}^{{\rm i} \t(z)} \,  I (z, z') \,
{\rm e}^{-{\rm i} \t(z')} ~;
\label{super-PDO1}
\ee
(ii) the equation  
\be
\z^A \cD_A \, I(z,z') 
= \z^A \Big( D_A +{\rm i} \, \G_A(z) \Big) I(z,z') =0~;
\label{super-PDO2}
\ee
(iii) the boundary condition 
\be 
I(z,z) ={\bf 1}~.
\label{super-PDO3}
\ee
These imply the important relation
\be
I(z,z') \, I(z', z) = {\bf 1}~.
\ee
We also have 
\be
\z^A \cD'_A \, I(z,z') 
= \z^A  \Big( D'_A \,I(z,z') 
 - {\rm i} \, I(z,z') \, \G_A(z') \Big) =0~.
\ee
Further, using the identity
\be
\z^B\, D_B \z^A = \z^A~, 
\ee
from (\ref{super-PDO2}) one deduces 
\be 
\z^{A_n} \ldots \z^{A_1} \, 
\cD_{A_1} \ldots \cD_{A_n} \, I(z,z') =0~.
\label{super-PDO4}
\ee
The latter leads to 
\be 
\cD_{( A_1} \ldots \cD_{A_n \} } \, I(z,z') \Big|_{z=z'} =0~,
\qquad n \geq 1~, 
\label{super-PDO5}
\ee
where $(\ldots \}$ means  graded symmetrization of $n$ indices 
(with a factor of $1/n!$).

Let $\J(z)$ be a superfield transforming in some representation of 
the gauge group, 
\be 
\J(z) ~\to ~ {\rm e}^{{\rm i} \t(z)} \, \J(z)~.
\ee
Then $U(z) \equiv I(z',z) \, \J(z)$ is gauge invariant with respect 
to $z$, and therefore we are in a position to apply the Taylor 
expansion (\ref{super-Taylor1}): 
\be
\J(z) = I(z,z') \,\sum_{n=0}^{\infty} {1 \over n!}\,
\z^{A_n} \ldots \z^{A_1} \, 
\cD_{A_1} \ldots \cD_{A_n} \, 
\Big(I(z',w)\, \J(w) \Big)\Big|_{w=z'}~. 
\ee
Because of the properties of $I(z,z')$, this is equivalent to 
the  covariant Taylor series 
\be 
\J(z) = I(z,z') \,\sum_{n=0}^{\infty} {1 \over n!}\,
\z^{A_n} \ldots \z^{A_1} \, 
\cD'_{A_1} \ldots \cD'_{A_n} \, \J(z') ~. 
\label{super-Taylor2}
\ee

The covariant Taylor expansion can be applied to $ \cD_B I(z,z') $
considered as a superfield at $z$, 
\be 
\cD_B I(z,z') = I(z,z') \,\sum_{n=0}^{\infty} {1 \over n!}\,
\z^{A_n} \ldots \z^{A_1} \, 
\cD_{A_1} \ldots \cD_{A_n} \cD_B\,I(w,z')\Big|_{w=z'}~. 
\ee
As is shown in the Appendix, this is equivalent to
\bea 
\cD_B I(z,z') &=& {\rm i} \, I(z,z') \,
\sum_{n=1}^{\infty} { 1  \over (n+1)!} \,
\Big\{
n \, \z^{A_n} 
\ldots \z^{A_1}  
\cD'_{A_1} \ldots \cD'_{A_{n-1} } F_{A_n \,B } (z') 
\label{super-PTO-der1} \\
&+& \hf  
(n-1)\, 
\z^{A_n} T_{A_n \,B}{}^C \,\z^{A_{n-1}} 
\ldots \z^{A_1}  
\cD'_{A_1} \ldots \cD'_{A_{n-2} } F_{A_{n-1} \,C } (z') \Big\}~.
\non 
\eea
Another form for this result is
\bea 
\cD_B I(z,z') &=& {\rm i} \,
\sum_{n=1}^{\infty} {(-1)^{n} \over (n+1)! } \, \Big\{
- \z^{A_n} 
\ldots \z^{A_1}  
\cD_{A_1} \ldots \cD_{A_{n-1} } F_{A_n \,B } (z) 
\label{super-PTO-der-mod} \\
&+& \hf  
(n-1)\,
\z^{A_n} T_{A_n \,B}{}^C \,\z^{A_{n-1}} 
\ldots \z^{A_1}  
\cD_{A_1} \ldots \cD_{A_{n-2} } F_{A_{n-1} \,C } (z) \Big\}
\, I(z,z')~.
\non 
\eea

${}$For the point $z^M(t)$ on the straight line 
$z^M (t) = (z-z')^M \, t+z'^M$, 
we introduce the notation
\be
\z^A(t) \equiv \z^A(z(t), z') =
t \,\z^A~,
\ee
with the two-point function $\z^A(z,z')$ as defined earlier. 
Then the results (\ref{super-PTO-der1}) 
and (\ref{super-PTO-der-mod})  can be expressed 
in integral form as
\bea
\cD_B \, I(z,z') &=& {\rm i} \, \int_0^1 {\rm d} t \,t \, I(z,z(t)) \,
\z^A \, F_{AB}(z(t)) \, I(z(t),z') \nonumber \\
&+& \frac{{\rm i}}{2} \, \int_0^1 {\rm d} t_1 \,t_1 \, I(z,z(t_1)) \,
\z^A \, T_{AB}{}^{C} \, \int_0^1 {\rm d} t_2 \, t_2 \, I(z(t_1),z(t_2t_1))
\nonumber \\
& \times & t_1 \, \z^D \, F_{DC}(z(t_2 t_1)) \, I(z(t_2 t_1),z')~.
\eea
The latter relation seems to be equivalent to the one derived in  
\cite{MM}. 

By analogy with the non-supersymmetric case, 
let us fix some point $z'$ in superspace
and consider the gauge transformation generated by 
\be
{\rm e}^{{\rm i} \t(z)}  = I (z', z) ~, \qquad 
{\rm e}^{{\rm i} \t(z')} =  {\bf 1}~. 
\ee
Applying this gauge transformation to $I(z,z')$, 
in accordance with (\ref{super-PDO1}), we end up with 
the superspace Fock-Schwinger gauge \cite{Ohr2} 
\be 
I(z,z') = {\bf 1} \quad \Longleftrightarrow \quad 
\z^A \, \G_A (z) =0~.
\ee
With this gauge choice, eq.  (\ref{super-PTO-der1}) 
tells us that the superconnection is
\bea 
\G_B (z) &=& 
\sum_{n=1}^{\infty} { 1  \over (n+1)!} \,
\Big\{
n \, \z^{A_n} 
\ldots \z^{A_1}  
\cD'_{A_1} \ldots \cD'_{A_{n-1} } F_{A_n \,B } (z')  \\
&+& \hf  
(n-1)\, 
\z^{A_n} T_{A_n \,B}{}^C \,\z^{A_{n-1}} 
\ldots \z^{A_1}  
\cD'_{A_1} \ldots \cD'_{A_{n-2} } F_{A_{n-1} \,C } (z') \Big\}~.
\non 
\eea

In the case of $\cN=1$ supersymmetry\footnote{Our $\cN=1$ notation 
and conventions correspond to \cite{BK}.}, 
the gauge covariant 
derivatives satisfy the following (anti)commutation relations:
\bea
& \{ \cD_a , \cD_\b \} 
= \{ {\bar \cD}_\ad , {\bar \cD}_\bd \} =0~, \qquad 
\{ \cD_\a , {\bar \cD}_\bd \} = - 2{\rm i} \, \cD_{\a \bd}~, \non \\
& [ \cD_\a , \cD_{\b \bd}] = 2 {\rm i} \ve_{\a \b}\,{\bar W}_\bd ~, 
\qquad 
[{\bar \cD}_\ad , \cD_{\b \bd}] = 2{\rm i} \ve_{\ad \bd}\,W_\b ~ , 
\non \\
& [ \cD_{\a \ad}, \cD_{\b \bd} ] = {\rm i} F_{\a \ad, \b\bd} 
= - \ve_{\a \b}\, {\bar \cD}_\ad {\bar W}_\bd 
-\ve_{\ad \bd} \,\cD_\a W_\b~, 
\label{N=1cov-der-al}
\eea
with the spinor field strengths $W_\a$ and ${\bar W}_\ad$ 
obeying the Bianchi  identities
\be
{\bar \cD}_\ad W_\a =0~, \qquad 
\cD^\a W_\a = {\bar \cD}_\ad {\bar W}^\ad~.
\ee
${}$For a covariantly constant gauge field, 
\be
\cD_a \, W_\b =0~, 
\label{constant SYM}
\ee 
the relation (\ref{super-PTO-der1}) is equivalent to the following: 
\bea
\cD_{\b \bd} I(z,z') &=& I(z,z') \,
\Big( -  \frac{{\rm i}}{4} \z^{\ad \a} F_{\a \ad, \b \bd}(z') 
-{\rm i} \, \z_{\b} \bar{W}_{\bd}(z') 
+ {\rm i} \, \bar{\z}_{\bd} W_{\b}(z')
\nonumber \\ 
&& \phantom{I(z,z') \,\Big( }
+\frac{2{\rm i}}{3} \, 
\bar{\z}_{\bd}
\z^{\a}
\cD_{\a}W_{\b}(z')
+ \frac{2{\rm i}}{3} \, \z_{\b} \bar{\z}^{\ad}
\bar{\cD}_{\ad} \bar{W}_{\bd}(z') \Big) \nonumber \\
&=& \Big( - \frac{{\rm i}}{4} \z^{\ad \a} F_{\a \ad, \b \bd}(z) 
-{\rm i} \, \z_{\b} \bar{W}_{\bd}(z) 
+ {\rm i} \, \bar{\z}_{\bd} 
W_{\b}(z)
\nonumber \\ 
&  & 
\phantom{\Big(}
-\frac{{\rm i}}{3} \, \bar{\z}_{\bd} \z^{\a} \cD_{\a}W_{\b}(z) 
- \frac{{\rm i}}{3} \, \z_{\b} \bar{\z}^{\ad}
\bar{\cD}_{\ad} \bar{W}_{\bd}(z) \Big) \, I(z,z')~;
\label{super-PTO-der2} \\
\cD_{\b} I(z,z') &=& I(z,z') \,
\Big( \frac{1}{12} \, \bar{\z}^{\bd} 
\z^{\a \ad}  F_{\a \ad, \b \bd}(z') 
- {\rm i} \, \z_{\b \bd} \Big\{
\frac{1}{2} 
\bar{W}^{\bd}(z') 
-\frac{1}{3}  
\bar{\z}^{\ad} \bar{\cD}_{\ad}
\bar{W}^{\bd}(z')  \Big\}
\nonumber \\ 
&  & 
+\frac13 \, \z_{\b} \bar{\z}_{\bd} \bar{W}^{\bd}(z')  
 +\frac13
\bar{\z}^2 \Big\{ 
W_{\b}(z') + \hf \z^{\a} \cD_{\a}
W_{\b}(z')
-\frac{1}{4}  \z_{\b} \cD^{\a} W_{\a}(z') \Big\}
\Big)
\nonumber \\
&=& \Big( \frac{1}{12} \, \bar{\z}^{\bd} 
\z^{\a \ad} \, F_{\a \ad , \b \bd}(z)
- \frac{{\rm i}}{2} \, \z_{\b \bd} 
\Big\{ 
\bar{W}^{\bd}(z) 
+\frac{1}{3}  \bar{\z}^{\ad} \bar{\cD}_{\ad}
\bar{W}^{\bd}(z) \Big\} 
+ \frac13 \, \z_{\b} \bar{\z}_{\bd} \bar{W}^{\bd}(z)  \nonumber \\ 
& & \phantom{ \Big( } +\frac13
\bar{\z}^2 \Big\{
W_{\b}(z) - \hf  \z^{\a} \cD_{\a} W_{\b}(z)
+\frac{1}{4}  \z_{\b} {\cD}^{\a} {W}_{\a}(z) \Big\}
\Big)\, I(z,z')~;
\label{super-PTO-der3} \\
{\bar \cD}_{\bd} I(z,z') &=& I(z,z') \,\Big( 
-\frac{1}{12} \, \z^{\b } 
\z^{\a \ad} \, F_{\a \ad , \b \bd}(z')
- {\rm i} \,\z_{\b \bd} \Big\{ 
\frac{1}{2}  W^{\b}(z') 
+\frac{1}{3}  \z^{\a} \cD_{\a} W^{\b}(z')  \Big\}
\nonumber \\ &  & 
- \frac13 \, \bar{\z}_{\bd} \z^{\b} W_{\b}(z')  
-\frac13 \z^2 \Big\{ 
\bar{W}_{\bd}(z')  -\hf \bar{\z}^{\ad} \bar{\cD}_{\ad}
\bar{W}_{\bd}(z')
+\frac{1}{4}  \bar{\z}_{\bd} {\bar \cD}^{\ad} 
{\bar W}_{\ad}(z') \Big\} 
\Big)
\nonumber \\
&=& \Big( - \frac{1}{12} \, \z^{\b} 
\z^{\a \ad}  F_{\a \ad , \b \bd}(z)
- \frac{{\rm i}}{2} \, \z_{\b \bd} \Big\{ W^{\b}(z) 
- \frac{1}{3}  \z^{\a} \cD_{\a} W^{\b}(z)  \Big\} 
- \frac13 \, \bar{\z}_{\bd} \z^{\b} W_{\b}(z)  \nonumber \\ 
& & - \frac13 \z^2 \Big\{ 
\bar{W}_{\bd}(z) +\hf  \bar{\z}^{\ad} \bar{\cD}_{\ad}
\bar{W}_{\bd}(z)
-\frac{1}{4}  \bar{\z}_{\bd} {\bar \cD}^{\ad} 
{\bar W}_{\ad}(z) \Big\}
\Big) \, I(z,z')~.
\label{super-PTO-der4} 
\eea

\sect{Green's functions in superspace}\label{super-Green's}

In the remaining sections of this paper, we will concentrate 
on $\cN=1$ supersymmetric Yang-Mills theories. 
A typical Green's function $G(z,z')$  
will be that of an unconstrained superfield transforming 
in some (real) representation of the gauge group. 
The Green's function satisfies the equation
\be 
\Box_z \, G(z,z') = -\d^8 (z-z')\, {\bf 1}~, \qquad 
\d^8 (z-z') = \d^4 (x-x') \, (\q-\q')^2
({\bar \q}-{\bar \q}')^2~, 
\label{super-Green1}
\ee
and the Feynman boundary conditions. 
Here the covariant d'Alembertian $\Box$ is \cite{GRS}
\bea 
\Box &=& \cD^a \cD_a -W^\a \cD_\a +{\bar W}_\ad {\bar \cD}^\ad \\
&=& -\frac{1}{8} \cD^\a {\bar \cD}^2 \cD_\a 
+{1 \over 16} \{ \cD^2 , {\bar \cD}^2 \} 
-W^\a \cD_\a -\hf  (\cD^\a W_\a) \non \\
&=& 
 -\frac{1}{8} {\bar \cD}_\ad \cD^2 {\bar \cD}^\ad 
+{1 \over 16} \{ \cD^2 , {\bar \cD}^2 \} 
+{\bar W}_\ad {\bar \cD}^\ad +\hf({\bar \cD}_\ad {\bar W}^\ad ) ~.
\non 
\eea
Our subsequent considerations require only a minor 
modification if the operator $\Box_z$ in (\ref{super-Green1}) 
is replaced with $\Box_z - \cP(z) $, with $\cP$ a local 
matrix function of the  gauge field. For simplicity, we set $\cP=0$.

In complete analogy with the non-supersymmetric case, 
we introduce the proper-time representation of $G(z,z')$.
The corresponding heat kernel,
\be
 K(z,z'|s) = {\rm e}^{ {\rm i} s \, (\Box \,+\,{\rm i} \,\ve)} 
 \, \d^8 (z-z') \, {\bf1}~,  \qquad \ve \to + 0~, 
\label{super-kernel1}
\ee
will be the main object of interest. 
Its gauge transformation law is 
\be
K(z,z'|s) ~\to ~ {\rm e}^{{\rm i} \t(z)} \, K(z,z'|s)\,
{\rm e}^{-{\rm i} \t(z')}~. 
\label{super-kernel2}
\ee
As in the non-supersymmetric case, 
it is useful to make use of the Fourier representation
\be
\d^8 (z-z') = \frac{1}{\p^4} \int {\rm d}^4 k
\int {\rm d}^2 \k \int {\rm d}^2 {\bar \k}\;
{\rm e}^{{\rm i} \, [k^a \z_a 
+ \k^\a \z_\a +{\bar \k}_\ad {\bar \z}^\ad]}~, 
\ee
where the supersymmetric interval $\z^A$ is
defined in eq. (\ref{super-interval}), and the
integration variables $k^a$ and 
$(\k^\a, {\bar \k}_\ad)$ are c-numbers and a-numbers, respectively.
In order to keep the gauge transformation law 
(\ref{super-kernel2}) manifest, we should actually represent
the full delta-function in the form
\be
\d^8 (z-z') \,{\bf 1}= \frac{1}{\p^4} \int {\rm d}^4 k
\int {\rm d}^2 \k \int {\rm d}^2 {\bar \k}\;
{\rm e}^{{\rm i} \, [k^a \z_a 
+ \k^\a \z_\a +{\bar \k}_\ad {\bar \z}^\ad]} \, I(z,z')~, 
\ee
with $I(z,z')$ the parallel dispacement propagator.
As a result, the heat kernel takes the form
\bea 
K(z,z'|s) &=&  \hat{K}(z,z'|s) \, I(z,z')~, \label{super-hatK}\\
\hat{K}(z,z'|s) & \equiv & 
\frac{1}{\p^4} \int {\rm d}^4 k
\int {\rm d}^2 \k \int {\rm d}^2 {\bar \k}\;
{\rm e}^{{\rm i} \, [k^a \z_a 
+ \k^\a \z_\a +{\bar \k}_\ad {\bar \z}^\ad]} \, 
{\rm e}^{{\rm i}s \, [X^a X_a  - W^\a X_\a 
+{\bar W}_\ad {\bar X}^\ad]}~, 
\non
\eea
where 
\be
X_a = \cD_a + {\rm i} k_a~, \quad 
X_\a = \cD_\a + {\rm i} \k_\a - k^a (\s_a)_{\a \ad } {\bar \z}^\ad~,
\quad 
{\bar X}^\ad = {\bar \cD}^\ad + {\rm i} {\bar \k}^\ad 
- k^a (\tilde{\s}_a)^{\ad \a} \z_\a~.
\ee
With respect to the gauge group, the operator $ \hat{K}(z,z'|s) $
transforms as 
\be
\hat{K}(z,z'|s) ~\to ~ {\rm e}^{{\rm i} \t(z)} \, \hat{K}(z,z'|s)\,
{\rm e}^{-{\rm i} \t(z)}~. 
\label{super-kernel3}
\ee

${}$For any gauge invariant superfield $\O(z)$ of compact support in 
space-time, we can establish the following operator identity
\be
\hat{K}(z,z'|s) \cdot \hat{\O} (z) = 
\hat{\O} (z') \cdot \hat{K}(z,z'|s)~, \qquad
\hat{\O}(z) \equiv \O(z) \,{\bf 1}~.
\label{super-master}
\ee
To prove this, one should  first introduce a generalized Fourier
representation of $\O$, 
\be
\O(z) =
\int {\rm d}^4 p
\int {\rm d}^2 \r \int {\rm d}^2 {\bar \r}\;
{\rm e}^{{\rm i} \, [p^a x_a 
+ \r^\a \q_\a +{\bar \r}_\ad {\bar \q}^\ad]}\,
\O(p,\r,{\bar \r})~, 
\ee
then push the ``plane wave'' $\exp ({\rm i} \,[ p^a x_a 
+ \r^\a \q_\a +{\bar \r}_\ad {\bar \q}^\ad ])$
through the operatorial exponential 
$\exp ({\rm i}s \, [X^a X_a  - W^\a X_\a 
+{\bar W}_\ad {\bar X}^\ad])$ in the 
integral representation (\ref{super-hatK}) of
$\hat{K}(z,z'|s) $, 
and finally introduce new integration variables 
by the rule
\be
k'^a = k^a +p^a~, \quad 
\k'^\a =\k^\a +\r^\a 
-{\rm i} \, p^a \,({\bar \q}' \tilde{\s}_a)^{ \a}~, \quad
{\bar \k}'_\ad =  {\bar \k}_\ad +{\bar \r}_\ad
- {\rm i} \, p^a \,(\q' \s_a)_{\ad}~.
\ee
Relation (\ref{super-master}) implies 
\be
\hat{K}(z,z'|s) = 
\mbox{{\bf :}} \,{\rm e}^{-\z^A \cD_A }\mbox{{\bf :}}
\, \hat{B} (z,z'|s) ~, 
\ee
where the two-point matrix $\hat{B} (z,z'|s) $ is a functional 
of the gauge  superfield such that 
\be
\hat{B}(z,z'|s) \cdot \hat{\O} (z) = 
\hat{\O} (z) \cdot \hat{B}(z,z'|s)~.
\ee
Therefore  $\hat{B} (z,z'|s) $ is not a differential operator.
By definition, the operator 
$\mbox{{\bf :}} \,{\rm e}^{-\z^A \cD_A }\mbox{{\bf :}}$
acts on a gauge covariant superfield $\J(z)$ 
(transforming
in some representation of the gauge group) as follows
\be
\mbox{{\bf :}} \,{\rm e}^{-\z^A \cD_A }\mbox{{\bf :}} \,
\J(z) = I(z,z') \, \J(z')~. 
\ee
If $\J(z)$ admits a covariant Taylor expansion, then 
\be
\mbox{{\bf :}} \,{\rm e}^{-\z^A \cD_A }\mbox{{\bf :}} \,
\J(z) = 
\sum_{n=0}^{\infty} {1 \over n!}\, (-1)^n\,
\z^{A_n} \ldots \z^{A_1} \, 
\cD_{A_1} \ldots \cD_{A_n} \, \J(z) 
\ee

Evaluation of the heat kernel (\ref{super-hatK}) can be carried out 
in a manner almost identical to that outlined 
at the end of sect. 2 for the non-supersymmetric case. 
The result is 
\be
K(z,z'|s) =-
\frac{\rm i}{( 4 \p  s)^{2}} \, 
{\rm e}^{{\rm i} \z^a \z_a/4s}\, F(z,z'|s)~, \qquad 
 F(z,z'|s) = \sum_{n=0}^{\infty} a_n (z,z') \,({\rm i}s)^n~,
\ee
where 
\bea
a_0 (z,z') &=& 
\mbox{{\bf :}} \,{\rm e}^{-\z^a \cD_a }\mbox{{\bf :}}
\, \d^4 (\q-\q') \, I(z,z') 
=  \d^4 (\q-\q') \,
\mbox{{\bf :}} \,{\rm e}^{-\z^a \cD_a }\mbox{{\bf :}}\, I(z,z') \non \\
&=&\d^4 (\q-\q') \,
\mbox{{\bf :}} \,{\rm e}^{-\z^A \cD_A }\mbox{{\bf :}}\, I(z,z') 
=\d^4 (\q-\q') \,I (z,z')~. 
\eea
Here $\d^4 (\q-\q') = (\q-\q')^2 ({\bar \q}-{\bar \q}')^2$,
and the operator 
$\mbox{{\bf :}} \,{\rm e}^{-\z^a \cD_a }\mbox{{\bf :}}$ is defined
similarly to (\ref{normal-exp}).

Along with the propagator $G(z,z')$ so far analysed, 
covariant supergraphs in $\cN=1$ super Yang-Mills theories
also involve (anti)chiral Green's functions\footnote{The 
proper-time representation for these Green's functions 
was developed in \cite{HSKS,McA}, see also
\cite{SY,BK85,BK}.}  
$G_\mp(z,z')$ associated with 
the covariantly chiral d'Alembertian
\bea 
\Box_+ &=& \cD^a \cD_a - W^\a \cD_\a -\hf \, (\cD^\a W_\a)~, 
\quad
\Box_+ \F = {1 \over 16} \, {\bar \cD}^2 \cD^2 \F ~, \quad 
{\bar \cD}_\ad \F =0~, 
\eea
and the covariantly antichiral d'Alembertian
\bea 
\Box_- &=& \cD^a \cD_a + {\bar W}_\ad {\bar \cD}^\ad 
+\hf \, ({\bar \cD}_\ad  {\bar W}^\ad)~, 
\quad
\Box_- {\bar \F} = {1 \over 16} \, \cD^2 {\bar \cD}^2  {\bar \F} ~, 
\quad  \cD_\a {\bar \F} =0~. 
\eea
Associated with the chiral Green's function $G_+(z,z')$
is the covariantly chiral heat kernel 
\bea
 K_+ (z,z'|s) &=& {\rm e}^{ {\rm i} s \, (\Box_+ \,+\,{\rm i} \,\ve)} 
 \, (-{1 \over 4} {\bar \cD}^2)\, \d^8 (z-z') \, {\bf1}~,  
\qquad \ve \to + 0~, 
\label{chiral-kernel1} \\
&& {\bar \cD}_\ad K_+ (z,z'|s) = {\bar \cD}'_\ad K_+ (z,z'|s) = 0~,
\non
\eea
and similarly in the antichiral case. 

In the case of an on-shell background gauge superfield, 
\be 
\cD^\a W_\a =0~,
\label{EoM}
\ee
which is often sufficient for applications, the (anti)chiral 
kernels $K_\mp (z,z'|s) $ do not require a separate treatment.
Indeed, we have
\be
\Box \,{\bar \cD}^2 = {\bar \cD}^2\, \Box ~, \qquad 
\Box \, {\bar \cD}^2 = \Box_+ \, {\bar \cD}^2 ~, 
\ee
and therefore
\be 
K_+ (z,z'|s) =   -{1 \over 4} {\bar \cD}^2  \, K (z,z'|s) ~.
\label{chiral-kernel2} 
\ee

The strategy explained in sect. 3 for computing 
multi-loop graphs is readily extended to the case 
of supergraphs by making use of the results of this section.

\sect{Superpropagator in a covariantly constant field}

In this section, we evaluate the heat kernel 
(\ref{super-kernel1})
in the the case of a covariantly constant gauge superfield
satisfying eq. (\ref{constant SYM}). 
Taken together with the Bianchi identities, 
this condition implies that the Yang-Mills supermultiplet
belongs to the Cartan subalgebra of the gauge group.
${}$For this background gauge superfield, 
the heat kernel (\ref{super-kernel1}) was originally 
computed by Ohrndorf \cite{Ohr1} using the Fock-Schwinger gauge 
in superspace \cite{Ohr2}. Here we derive, 
for the first time, this exact 
solution in a manifestly gauge covariant way.

It follows from (\ref{N=1cov-der-al}) that
\be
\cD_a W_{\b} = 0 \quad \Longrightarrow \quad
[\cD_a, W^{\b} \cD_{\b} - \bar{W}_{\bd}
{\bar \cD}^{\bd}] = 0~.
\ee
This identity
allows a convenient factorization of the kernel in the form
\be
K(z,z'|s) = U(s) \, {\rm  e}^{{\rm i} s \, \cD^a \cD_a} \, 
\delta^8( z- z') \,{\bf 1}~, \qquad 
U(s) = {\rm e}^{- {\rm i} s (W^{\a} \cD_{\a} 
- \bar{W}_{\ad} {\bar \cD}^{\ad})}~.
\ee
As a result, it proves efficient to use a Fourier transformation 
of only the bosonic part of the superspace delta function,
\be
\delta^8(z-z') = \int \frac{{\rm d}^4k}{(2 \pi)^4} \, 
{\rm e}^{{\rm i} k^a \z_a} \,
\z^2 \bar{\z}^2~,
\ee
where $\z^A$ is defined in eq. (\ref{super-interval}).

The kernel of interest (\ref{super-kernel1})
is then obtained by the action of the operator $U(s)$ on the
simpler kernel
\be
\tilde{K}(z,z'|s)  =  \int \frac{{\rm d}^4k}{(2 \pi)^4} \, 
{\rm e}^{{\rm i} k^a \z_a} \,
{\rm e}^{{\rm i} s  (\cD + ik)^2}\,  \z^2 \, \bar{\z}^2
\, I(z,z')  ~,
\label{SSK2}
\ee
where $I(z,z')$ is the parallel displacement operator.
The operator $U(s)$ acts to ``shift''
the $\z^A$ dependence of $\tilde{K}(z,z'|s)$.
With the notation $N_{\a}{}^{\b} = \cD_{\a} W^{\b}$,
\bea
U(s) \, W^{\a} \, U(-s) & \equiv &  W^{\a}(s) 
= W^{\b} ( {\rm e}^{- {\rm i} s N} )_{\b}\,{}^{\a}~, 
\nonumber \\
U(s) \, \z^{\a} \, U(-s) & \equiv &  \z^{\a}(s) = \z^{\a} + W^{\b} \,
( ({{\rm e}^{-{\rm i}s N} -1})\, {N}^{-1} )_{\b}\,{}^{\a}~, \nonumber \\
U(s) \, \z_{\a \ad} \, U(-s) & \equiv &  \z_{\a \ad}(s) = \z_{\a \ad} 
-2 \int_0^{s} {\rm d}t \, \Big( W_{\a}(t) \bar{\z}_{\ad}(t) 
+ \z_{\a}(t)\bar{W}_{\ad}(t) \Big)~.
\label{zeta(s)}
\eea

The kernel $\tilde{K}(z,z'|s)$ satisifies the equation
\be
\left ( {\rm i}\, \frac{\rm d}{{\rm d}s} +  \cD^a \cD_a \right)
\tilde{K}(z,z'|s) = 0
\label{SSKde}
\ee
with the boundary condition $\lim_{s \rightarrow 0} \tilde{K}(z,z'|s) =
\delta^8(z-z')\,{\bf 1}$.
The identity
\bea
0 &=& \int
{{\rm d}^4k} \,
\frac{\partial}{\partial k^a} \Big(
{\rm e}^{{\rm i} k^b \z_b} \, {\rm e}^{{\rm i} s
(\cD + ik)^2}\,  \z^2
\, \bar{\z}^2 \, I(z,z')
\Big) \nonumber \\
&=& {\rm i} \, \z_a \, \tilde{K}(z,z'|s) - 2 \, s \int
{{\rm d}^4k} \, {\rm e}^{{\rm i} k^b \z_b}  \int_0^1 {\rm d}t \,
  \, {\rm e}^{{\rm i} s t
(\cD + ik)^2}\, (\cD_a + {\rm i} k_a) \, {\rm e}^{- {\rm i} st
(\cD + ik)^2}\, \nonumber \\
& & \phantom{{\rm i} \, \z_a \, \tilde{K}(z,z'|s) }
\times  
{\rm e}^{{\rm i} s 
(\cD + ik)^2}\, 
\z^2
\, \bar{\z}^2 \, I(z,z')
\eea
can be used to show that
\be
\cD_a \, \tilde{K}(z,z'|s) = {\rm i}\,
\left( \frac{F}{{\rm e}^{2  s F} - 1} \right)_{ab} \, \z^b \,
\tilde{K}(z,z'|s)~.
\label{DaK}
\ee
Differentiating (\ref{DaK}) again allows the right hand side of
(\ref{SSKde}) to be expressed in terms of $\tilde{K},$ and
equation (\ref{SSKde}) can then be integrated in the form
\be
\tilde{K}(z,z'|s) = -\frac{\rm i}{16 \pi^2} \, 
\det\left( \frac{2 \,F}{{\rm e}^{ 2  s F} -1}\right)^{\frac12} 
\, {\rm e}^{ \frac{{\rm i}}{4} \z^a (F \coth (s F))_{ab} \z^b}
\, \z^2 \, \bar{\z}^2 \,C(z,z')
~,
\label{Ksol}
\ee
where the determinant is computed with respect 
to the Lorentz indices.
Here, $C(z,z')$ is an integration constant which
must transform appropriately under the gauge group 
and satisfy the boundary condition $C(z,z) = {\bf 1}$.
Substituting (\ref{Ksol}) into (\ref{DaK}) yields the further condition
\be
  \z^2 \, \bar{\z}^2 \, \cD_a C(z,z') = -\, \frac{{\rm i}}{2} \,  \z^2 \,
\bar{\z}^2 \,  F_{ab} \z^b \, C(z,z')~,
\label{DaI}
\ee
as the action of $\cD_a$
on the expression 
$\exp [ {  \frac{{\rm i}}{4} \z^a (F \coth ( s F))_{ab} \z^b}]$ 
produces the symmetric part of 
$\left( \frac{{\rm i} \,F}{{\rm e}^{ 2  s F} - 1} \right)$ on the
right hand side of (\ref{DaK}), but not the antisymmetric part 
$(- \frac{{\rm i}}{2} F)$.
In accordance with (\ref{super-PTO-der2}), we can choose 
$C(z,z') =  I(z,z')$. As a result we arrive at the kernel
\bea
K(z,z'|s) &=& -\frac{\rm i}{(4 \pi s)^2} \, 
\det\left( \frac{2\, s \,F}{{\rm e}^{ 2  s F} -1}\right)^{\frac12} 
\, {\rm e}^{ \frac{{\rm i}}{4} \z^a (s) 
(F \coth ( s F))_{ab} \z^b (s)} \, \non \\
& \times &\z^2(s) \, \bar{\z}^2 (s)\,
U(s)\, I(z,z')~, 
\label{exactsuper-kernel}
\eea
with $\z^A(s) $ defined in (\ref{zeta(s)}).
The final ingredient is
\be
U(s) \, I(z,z') = \exp \,\Big\{
\int_0^s {\rm d}t \, \Xi(\z(t), W(t),
\bar{W}(t)) \Big\} \, I(z,z')~,
\ee
where
\bea
\Xi(\z(s), W(s),  \bar{W}(s)) &=& U(s) \, \Xi(\z, W,
\bar{W}) \, U(-s)~, \non \\
\Xi (\z, W
\bar{W}) &=& \frac{1}{12} \, \z^{\ad \a}\Big( W^{\b}\bar{\z}^{\bd} 
- \z^{\b} \bar{W}^{\bd}\Big) \Big(\ve_{\b \a} \, 
\bar{\cD}_{\bd} {\bar W}_{\ad} 
- \ve_{\bd \ad} \, \cD_{\b}W_{\a}\Big) \nonumber \\
& - & \, \frac{2 {\rm i}}{3}\, \z W  \, \bar{\z} \bar{W} 
- {{\rm i} \over 3} \, \z^2 \, \Big( \bar{W}^2
- \frac{1}{4} \bar{\z}
\bar{\cD} \,\bar{W}^2 
+ \frac{1}{4} \bar{\z} \bar{W}\, {\bar \cD} {\bar W} \Big) \nonumber \\
&-& {{\rm i} \over 3}\, \bar{\z}^2 \, \Big(  W^2 - \frac{1}{4} \z
\cD \,W^2
  + \frac{1}{4} \z W \, \cD W \Big)~.
\eea

If the background gauge superfield is further constrained
to satisfy the equation of motion (\ref{EoM}), then 
from (\ref{exactsuper-kernel}) we can immediately derive the chiral 
kernel $K_+(z,z'|s)$ using the prescription (\ref{chiral-kernel2}).
The result for $K_+(z,z'|s)$ can be shown to agree with the 
exact superpropagator computed in \cite{BK85}
(see also a related work \cite{SY}), although the final relation 
in (the English translation of) \cite{BK85} contains a misprint.

In conclusion, we would like to comment on the relationship 
between $K_+(z,z'|s)$ and the chiral propagator used 
in \cite{DGLVZ} for a calculation in the context of 
the matrix model/gauge theory correspondence \cite{DV}.
The authors of  \cite{DGLVZ} evaluate chiral loop diagrams  
in an on-shell covariantly constant SYM background,
as specified by eqs. (\ref{constant SYM}) and (\ref{EoM}),
which is then further simplified 
by formally setting ${\bar W}_\ad =0$ and $F_{ab} = 0$
while keeping $W_\a \neq 0$.
For such a background, in the chiral representation
$K_+(z,z'|s)$ is 
\be
K_+(z,z'|s) = -\frac{\rm i}{(4 \pi s)^2} \, 
{\rm e}^{ {\rm i}(x-x')^2 /4s } \, 
(\q -\q' -{\rm i} s \,W)^2
\, I(z,z')~. 
\ee
This is the heat kernel corresponding to the chiral 
propagator used in \cite{DGLVZ}.

\vskip.5cm

\noindent
{\bf Acknowledgements.}
This work is supported in part by the Australian Research
Council, the Australian Academy of Science as well as by UWA research 
grants.

\begin{appendix}

\sect{Technical lemma}
This appendix is devoted to a derivation of eq. (\ref{super-PTO-der1}).
It is in fact sufficient to prove the following relation
\bea
&(n+1) \, \z^{A_n} \ldots \z^{A_1} \cD_{A_1} \ldots 
 \cD_{A_n} \cD_B I(z,z')\Big|_{z=z'} 
= n \, \z^{A_n} \ldots \z^{A_1} \cD_{A_1} \ldots 
\cD_{A_{n-1}} F_{A_n \,B} \non \\
& + \hf (n-1)\, 
\z^{A_n} T_{A_n \,B}{}^C \,\z^{A_{n-1}} 
\ldots \z^{A_1}  
\cD_{A_1} \ldots \cD_{A_{n-2} } F_{A_{n-1} \,B }~,
\label{A1}
\eea
with $n$ a positive integer.

We start with an obvious identity
\bea
 (n &+& 1) \, \z^{A_n} \ldots \z^{A_1} \cD_{( A_1} \ldots 
 \cD_{A_n} \cD_{ B\} } 
= \z^{A_n} \ldots \z^{A_1} \cD_{A_1} \ldots 
 \cD_{A_n} \cD_B \non \\
& + & \sum_{i=1}^{n} (-1)^{ B (A_i + \ldots + A_n) }
\z^{A_n} \ldots \z^{A_1} \cD_{ A_1} \ldots \cD_{A_{i-1}}
\cD_B \cD_{A_i} \ldots  \cD_{A_n}~,
\eea
and make use of eq. (\ref{super-PDO5}), rewritten as 
\be 
\cD_{( A_1} \ldots \cD_{A_n} \cD_{B \} } \, I(z,z') \Big|_{z=z'} =0~.
\ee
We thus have 
\bea 
0 &=& \z^{A_n} \ldots \z^{A_1} \cD_{A_1} \ldots 
 \cD_{A_n} \cD_B \, I(z,z') \Big|_{z=z'} \non \\
& + & \sum_{i=1}^{n} (-1)^{ B (A_i + \ldots + A_n) }
\z^{A_n} \ldots \z^{A_1} \cD_{ A_1} \ldots \cD_{A_{i-1}}
\cD_B \cD_{A_i} \ldots  \cD_{A_n} \, I(z,z') \Big|_{z=z'}~. 
\label{A4}
\eea
The next natural step is to represent 
\bea
&& (-1)^{ B (A_i + \ldots + A_n) }
\z^{A_n} \ldots \z^{A_1} \cD_{ A_1} \ldots \cD_{A_{i-1}}
\cD_B \cD_{A_i} \ldots  \cD_{A_n} \non \\
=&-&
(-1)^{ B (A_{i+1} + \ldots + A_n) }
\z^{A_n} \ldots \z^{A_1} \cD_{ A_1} \ldots \cD_{A_{i-1}}
[\cD_{A_i} , \cD_B \} \cD_{A_{i+1}} \ldots  \cD_{A_n} \non \\
&+& (-1)^{ B (A_{i+1} + \ldots + A_n) }
\z^{A_n} \ldots \z^{A_1} \cD_{ A_1} \ldots \cD_{A_i}
\cD_B \cD_{A_{i+1}} \ldots  \cD_{A_n}
\eea 
and make use of the covariant derivative algebra 
(\ref{CDA}), along with the observation 
\bea
&(-1)^{ B (A_{i+1} + \ldots + A_n) }
\z^{A_n} \ldots \z^{A_1} \cD_{ A_1} \ldots \cD_{A_{i-1}}
F_{A_i \, B} \cD_{A_{i+1}}\ldots  \cD_{A_n}\, 
I(z,z') \Big|_{z=z'} \non \\
& \qquad = \left\{ 
\begin{array}{cl}
0~, & \qquad \quad i <n ~; \\
\z^{A_n} \ldots \z^{A_1} \cD_{ A_1} \ldots \cD_{A_{n-1}}\,
F_{A_n \,B}~, & \qquad \quad i=n~.
\end{array} 
\right. 
\eea
Repeating this procedure, each  contribution to the second terms in  
(\ref{A4}) can be reduced to the first term  
plus additional terms involving graded commutators of covariant derivatives.
Since the torsion $T_{AB}{}^C$ in (\ref{CDA}) is constant, 
we then obtain
\bea
&&(n+1) \, \z^{A_n} \ldots \z^{A_1} \cD_{A_1} \ldots 
 \cD_{A_n} \cD_B I(z,z')\Big|_{z=z'} \non \\
&=&  
\sum_{i=1}^{n} i\, (-1)^{ C (A_{i+1} + \ldots + A_n) }
\z^{A_i} T_{A_i\,B}{}^C
\z^{A_n} \ldots 
\underbrace{1}_{i} 
\ldots \z^{A_1} 
\cD_{ A_1} \ldots 
\underbrace{\cD_C}_{i} 
\ldots  \cD_{A_n} \, I(z,z') \Big|_{z=z'} \non \\
&& + n {\rm i} \, 
\z^{A_n} \ldots \z^{A_1} \cD_{ A_1} \ldots \cD_{A_{n-1}}\,
F_{A_n \,B}~.
\eea
${}$For the first term in the right hand side, 
we can again apply  the previous procedure, 
and this now simplifies since 
\be 
T_{AB}{}^C\, [\cD_C , \cD_D \} = 
(-1)^{C}\, T_{AB}{}^C\, [\cD_C , \cD_D \} = 
{\rm i}\, T_{AB}{}^C\, F_{CD}~.
\ee
After some algebra, one then arrives at (\ref{A1}).

\end{appendix}

\end{document}